From sabh@202.38.8.5 Fri Apr 23 05:37:08 1999
Date: Tue, 20 Apr 1999 06:57:02 +0800
From: Sa Ben-Hao <sabh@202.38.8.5>
To: caix@iopp.ccnu.edu.cn

  [ Part 2: "Attached Text" ]

\documentstyle[preprint,aps]{revtex}
\begin{document}
\tighten

\title{Enhancement of singly and multiply strangeness in p-Pb and
Pb-Pb collisions at 158A GeV/c}
\vspace{0.1in}
\author{
  Sa Ben-Hao$^{1,2,5}$, Wang Xiao-Rong$^3$, Tai An$^{1,4}$, Zhou Dai-Cui$^3$,
   and Cai Xu$^3$ }

\address{
$^1$  CCAST (World Lab.), P. O. Box 8730 Beijing, 100080
China. \\
$^2$  China Institute of Atomic Energy, P. O. Box 275 (18),
Beijing, 102413 China. \footnotemark \\
$^3$  Institute of particle Physics, Huazhong normal University,
 Wuhan, 430079 China.\\
$^4$  Dept. of Physics and Astronomy, Univ. of California at
 Los Angeles, Los Angeles, CA 90025\\
$^5$  Institute of Theoretical Physics, Academia Sinica,
Beijing, 100080 China.
\footnotetext{mailing address. E-mail: sabh@iris.ciae.ac.cn}
}

\maketitle
\begin{abstract}
The idea that the reduction of the strange quark suppression in string
fragmentation leads to the enhancement of strange particle yield in
nucleus-nucleus collisions is applied to study the singly and multiply
strange particle production in p-Pb and Pb-Pb collisions at 158A GeV/c.
In this mechanism the strange quark suppression factor is related to the
effective string tension, which increases in turn with the increase of
the energy, of the centrality and of the mass of colliding system. The
WA97 observation that the strange particle enhancement increases with
the increasing of centrality and of strange quark content in multiply
strange particles in Pb-Pb collisions with respect to p-Pb collisions
was accounted reasonably.

\end{abstract}

\section{Introduction}
Strangeness as a possible signature of the phase transition from a
hadronic state to a QGP state was put forward about 16 years ago
\cite{rafelski1}. It was based on the prediction that the
production of strange quark pairs would be enhanced as a result of
the approximate chiral symmetry restoration in a QGP state in
comparison with a hadronic state. The strangeness enhancement in
pA and AA collisions with respect to the superposition of
nucleon-nucleon collisions has been investigated and confirmed by
many experimental groups \cite{wa85,na351,na36,wa97,naga}.
However, alternative explanations exist, they are based on the `conventional' 
physics in the hadronic regime, like rescattering, string-string
interaction, etc. \cite{satai,venus,rqmd}. The first
detailed theoretical study of strangeness production can be found
in \cite{rafelski2}, where the enhanced relative yield of strange
and multi-strange particles in nucleus-nucleus collisions with
respect to proton-nucleus interactions has been suggested as a
sensitive signature of a QGP.

We have done a series of studies in recent years investigating
strangeness enhancement with a hadron and string scenario
\cite{satai,satai2,satai3,taisa1,taisa2}, from which a Monte-Carlo
event generator, LUCIAE, was developed \cite{luciae}. Those
studies indicate that including rescattering of the final state
hadrons is still not enough to reproduce the NA35 \cite{na351} data of
strange particle production. To reproduce the NA35 data needs to
rely further on the mechanism of reduction of the strange quark
suppression in string fragmentation, which contributes to the enhancement of
strange particle yield in nucleus-nucleus collisions with respect
to the superposition of the nucleon-nucleon collisions
\cite{satai2,satai3,taisa1,taisa2}. Similarly, in order to
reproduce the NA35 data, the RQMD generator, equipped with
rescattering though, has to resort to the colour rope mechanism
\cite{rqmd}. In this picture it is assumed that the neighboring
interacting strings might form a string cluster called colour rope
in pA and AA collisions. The colour rope then fragments in a
collective way and tends to enhance the production of the strange
quark pairs from the colour field of strings through the increase
of the effective string tension.

It has been known for years that the strange quark suppression
factor ($\lambda$ hereafter), i.e., the suppression of s quark
pair production in the color field with respect to u or d pair
production, in hadron-hadron collisions is not a constant, but
energy-dependent, increasing from a value of 0.2 at the ISR
energies to about 0.3 at the top of the SPS energies \cite{kapa}.
In \cite{taisa1} we proposed a mechanism to investigate the energy
dependence of $\lambda$ in hh collisions by relating the effective
string tension to the production of hard gluon jets (mini-jets). A
parameterization form was then obtained, which reproduces the
energy dependence of $\lambda$ in hh collisions reasonably well
\cite{taisa1}. When the same mechanism is used in the study of 
pA and AA collisions it is found that $\lambda$ would increase
with the increase of energy, mass and centrality of a colliding system
as a result of mini-jet(gluon) production stemming from the
string-string interaction. Our model reproduced nicely the data of
strange particle production in hh \cite{taisa1}, pA, and AA
\cite{satai3,taisa2} collisions.

In this work we use above ideas \cite{taisa1,taisa2} to study the
recently published WA97 data of the enhanced production of singly
and multiply strange particles in p-Pb and Pb-pb collisions at
158A GeV/c. The study indicates that the WA97 data, which 
revealed that the enhancement of strange particle yield
increases with the increasing of centrality and of s quark content
in multiply strange particles in Pb-Pb collisions with respect
to p-Pb collisions, could be explained in a hadron-string model
except for $\Omega$ yield in the Pb-Pb data.

\section{Brief review of the LUCIAE model}
LUCIAE model is developed based on the FRITIOF model
\cite{fritiof}. FRITIOF is a string model, which started from the
modeling of inelastic hadron-hadron collisions and it has been
successful in describing many experimental data from the low
energies at the ISR-regime all the way to the SPS energies
\cite{B.N,H.P1}. In this model a hadron is assumed to behave like
a massless relativistic string. A hadron-hadron collision is
pictured as the multi-scattering of the partons inside the two
colliding hadrons. In FRITIOF, during the collision two hadrons
are excited due to longitudinal momentum transfers and/or a
Rutherford Parton Scattering (RPS). The highly excited states will
emit bremsstrahlung gluons according to the soft radiation model.
They are afterwards treated as excitations i.e. the Lund Strings
and allowed to decay into final state hadrons according to the
Lund fragmentation scheme.

The FRITIOF model has been extended to also describe
hadron-nucleus and nucleus-nucleus collisions by assuming that the
reactions are superposition of binary hadron-hadron collisions in
which the geometry of the nucleus plays an important role because
the nuclei should then behave as a ``frozen'' bag of nucleons.
However in the relativistic nucleus-nucleus collisions there are
generally many excited strings formed close by each other during a
collision. Thus in the LUCIAE model a Firecracker model
\cite{fire} is proposed to deal with the string-string collective
interaction. In the Firecracker model it is assumed that several
string from a relativistic heavy ion reaction will form a cluster
and then the strings inside such a cluster will interact in a
collective way. We assume that the groups of neighbouring strings
in a cluster may form interacting quantum states so that both the
emission of gluonic bremsstrahlung as well as the fragmentation
properties can be affected by the large common energy density.

In relativistic nucleus-nucleus collision there are generally a
lot of hadrons produced, however, FRITIOF does not include the
final state interactions. Thus in LUCIAE a rescattering model
\cite{satai} is devised to consider the reinteraction of the
produced hadrons with each other and with the surrounding cold
spectator matter. The distributions of the final state hadrons
will be affected by the rescattering process. We refer to the
Refs. \cite{satai,luciae} for the details and we just give here
the list of the reactions involving in LUCIAE, which are cataloged
into
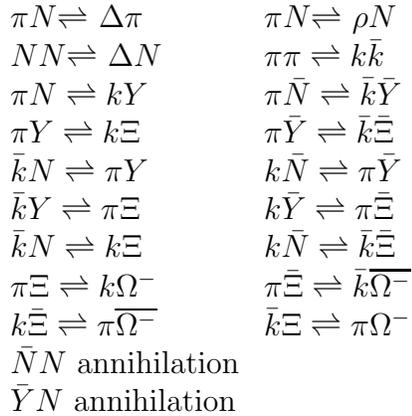
\begin{tabbing}
ttttttttttttttt\=ttttttttttttttt\=tttttt\=tttttttttttttttt\=  \kill
\>$\pi$$N$$\rightleftharpoons$ $\Delta$$\pi$
\> \>$\pi$$N$$\rightleftharpoons$ $\rho$$N$\\
\> $N$$N$$\rightleftharpoons$ $\Delta$$N$
\> \>$\pi\pi \rightleftharpoons k\bar{k}$\\
\>$\pi N \rightleftharpoons kY$
\> \>$\pi\bar{N} \rightleftharpoons  \bar{k}\bar{Y}$\\
\>$\pi Y  \rightleftharpoons k\Xi$
\> \>$\pi\bar{Y}  \rightleftharpoons  \bar{k}\bar{\Xi}$\\
\>$\bar{k}N  \rightleftharpoons  \pi Y$
\> \>$k\bar{N}  \rightleftharpoons  \pi\bar{Y}$\\
\>$\bar{k}Y  \rightleftharpoons  \pi\Xi$
\> \>$k\bar{Y}  \rightleftharpoons  \pi\bar{\Xi}$\\
\>$\bar{k}N  \rightleftharpoons  k\Xi$
\> \>$k\bar{N}  \rightleftharpoons  \bar{k}\bar{\Xi}$\\
\>$\pi\Xi \rightleftharpoons k\Omega^- $
\> \>$\pi\bar{\Xi} \rightleftharpoons  \bar{k}\overline{\Omega^-}$\\
\>$k\bar{\Xi} \rightleftharpoons \pi\overline{\Omega^-}$
\> \>$\bar{k}\Xi \rightleftharpoons \pi\Omega^-$\\
\>$\bar{N}N$ annihilation\\
\>$\bar{Y}N$ annihilation\\
\end{tabbing}
where $Y$ refers to the $\Lambda$ or $\Sigma$ and $\Xi$ refers to the
$\Xi^-$ or $\Xi^0$. There are 364 reactions involved altogether.

In addition, the reduction mechanism of s quark suppression, i. e., the s 
quark suppression factor increasing with energy, centrality, and mass of the 
colliding system, which is linked to string tension, is included in
LUCIAE via the parameterized formulas \cite{taisa1,taisa2}
\begin{equation}
\kappa_{eff}=\kappa_{0} (1-\xi)^{-\alpha},
\label{f2}
\end{equation}
where $\kappa_{0}$ is the string tension of the pure $q\bar{q}$ string,
$\alpha$ is a parameter $\sim$ 3, and $\xi$ ($\leq$ 1) is calculated by
\begin{equation}
\xi =\frac{\ln(\frac{k_{\perp max}^2}{s_{0}})}{\ln (\frac{s}{s_{0}}) +
\sum_{j=2}^{n-1} \ln (\frac{k_{\perp j}^2}{s_{0}})},
\label{f3}
\end{equation}
which represents the scale that a multigluon string is deviated from a pure
$q\bar{q}$ string.

The s quark suppression factor, $\lambda$, of two string states can thus 
be calculated by
\begin{equation}
\lambda_{2} = \lambda_{1}^ {\frac{\kappa_{eff1}}{\kappa_{eff2}}},
\label{f1}
\end{equation}
where $\kappa_{eff}$ refers to the effective string tension of
a multigluon string. Since $\lambda$ is always less than one, above
equation indicates the larger effective string tension the more
reduction of s quark suppression. The effective string tension is
then relevant to the hard gluon kinks (mini-(gluon) jets) created
on the string.

It should be mentioned that the LUCIAE (FRITIOF) event generator
runs together with JETSET routine. In JETSET routine there are
model parameters PARJ(2) (i.e., $\lambda$) and PARJ(3). PARJ(3) is
the extra suppression of strange diquark production compared to
the normal suppression of strange quark pair. Both PARJ(2) and
PARJ(3) are responsible for the s quark (diquark) suppression and
related to the effective string tension (the relation of Eq. (3)
holds true for PARJ(3) as for $\lambda$). Besides $\lambda$ and PARJ(3) there 
is PARJ(1), which stands for the suppression of diquark-antidiquark
pair production in the color field in comparison with the
quark-antiquark pair production and is related to the effective
string tension as well. The mechanism mentioned above is performed
via these parameters in program. How these three parameters affect
the multiplicity distribution of final state particles can be
found in \cite{satai2,satai3}.

\section{Results and discussions}

In Table 1 is given the results of the JETSET parameters PARJ(1),
PARJ(2) (i.e., $\lambda$), and PARJ(3) varying with the centrality
and the size of collision system in p-Pb and Pb-Pb collisions
at 158A GeV/c. That seems quite reasonable.

Fig. 1a shows the calculated $\Lambda+\bar{\Lambda}$, $\Xi^-+
\overline{\Xi^-}$, and $\Omega^-+\overline{\Omega^-}$ yields per event
($|y-y_{cm}| \leq$ 0.5 and p$_T$ $\geq$ 0 GeV/c) as a function of the 
number of participant in minimum bias p-Pb collisions and in central 
(b=2) Pb-Pb collisions at 158A GeV/c (open labels) comparing
with WA97 data (full labels) \cite{wa97}. The corresponding
results in Pb-Pb collisions after recaling each yield according
to its value in p-Pb are given in Fig. 1b. One knows from Fig. 1a that 
the agreement between theory and experiment is quite well
for $\Lambda+\bar{\Lambda}$ and $\Xi^-+\overline{\Xi^-}$, however,
for $\Omega^-+\overline{\Omega^-}$ the theoretical results are lower
than experiments. That should be study further both theoretically
and experimentally. In Fig. 1b the theoretical results of
$\Omega^-+\overline{\Omega^-}$ are also lower than experiments,
however, the trend of the strangeness enhancement increasing with
increase of the centrality and of the s quark content in strange
particles is reproduced quite well.

In Fig. 2 and 3 are given, respectively, the calculated m$_T$
spectra ($|y-y_ {cm}| \leq$ 0.5) of $\Lambda$, $\bar{\Lambda}$,
$\Xi^-$, $\overline{\Xi^-}$ and $\Omega^-+\overline{\Omega^-}$ in
p-Pb and Pb-Pb collisions at 158A GeV/c (open labels). The
corresponding full labels in those figures are the corresponding WA97
data \cite{wa97}. One sees from figure 2 that the agreement between
theory and experiment is reasonably good, except that the fluctuation
in theoretical results of $\Omega^-+\overline{\Omega^-}$ m$_T$
spectrum has to be improved. However, the situations in figure 3
is much better, i.e., the agreement between theory and experiment
is reasonably good.

In summary, we have used a hadron and string cascade model,
LUCIAE, to investigate the WA97 data of the strangeness
enhancement increasing with the increase of the centrality and of
the s quark content in strange particles. Relying on the mechanism of
the reduction of s quark suppression in string fragmentation leads
to the enhancement of strange particle yield in nucleus-nucleus
collisions the WA97 data could be reproduced nicely except $\Omega$ yield
in Pb+Pb collisions, which need to be studied further.

\section{ACKNOWLEDGMENTS}
We would like to thank T. Sj\"{o}strand for detailed instructions
of using PYTHIA. This work was supported by national Natural
Science Foundation of China and Nuclear Industry Foundation of
China.

\newpage

\begin{center}Figure Captions\end{center}
\begin{quotation}
Fig. 1 a) The calculated $\Lambda+\bar{\Lambda}$, $\Xi^-+
\overline{\Xi^-}$, and $\Omega^-+\overline{\Omega^-}$ yields per event
($|y-y_{cm}| \leq$ 0.5 and p$_T$ $\geq$ 0 GeV/c) as a function
of the number of participant in p-Pb and Pb-Pb collisions at
158A GeV/c (open labels) comparing with WA97 data (full labels)
\cite{wa97}. b) The calculated $\Lambda+\bar{\Lambda}$, $\Xi^-+
\overline{\Xi^-}$, and $\Omega^-+\overline{\Omega^-}$ yields per event in
Pb-Pb expressed in units of the corresponding yields in p-Pb
as a function of the number of participant in Pb-Pb (open
labels), the full labels are the corresponding WA97 data
\cite{wa97}.

Fig. 2 The calculated m$_T$ spectra ($|y-y_{cm}| \leq$ 0.5) of
$\lambda$, $\bar{\lambda}$, $\Xi^-$, $\overline{\Xi^-}$ and $\Omega^-+
\overline{\Omega^-}$ in p-Pb collisions at 158 GeV/c (open labels)
comparing with WA97 data (full labels) \cite{wa97}.

Fig. 3 The calculated m$_T$ spectra ($|y-y_{cm}| \leq$ 0.5) of
$\lambda$, $\bar{\lambda}$, $\Xi^-$, $\overline{\Xi^-}$ and $\Omega^-+
\overline{\Omega^-}$ in Pb-Pb collisions at 158A GeV/c (open labels)
comparing with WA97 data (full labels) \cite{wa97}.
\end{quotation}
\newpage
%\vspace{0.5cm}

\begin{table}
\caption{The values of the JETSET parameters in
p-Pb and Pb-Pb collisions at 158A GeV/c}
\label{Table:table1}
\begin{tabular}{ccccccc}
&\multicolumn{2}{c}{p+Pb} &\multicolumn{4}{c}{Pb+Pb}\\
&b=0 & b=5 & b=0 &b=2 &b=4&b=6\\ \hline
PARJ(1) &0.0852& 0.07734 &0.1180 &0.1171 &0.1142 &0.1105\\
PARJ(2) &0.2686& 0.2577  &0.3146 &0.3134 &0.3096 &0.3048\\
PARJ(3) &0.3652& 0.3549  &0.4105 &0.4093 &0.4056 &0.4010\\
\end{tabular}
\end{table}
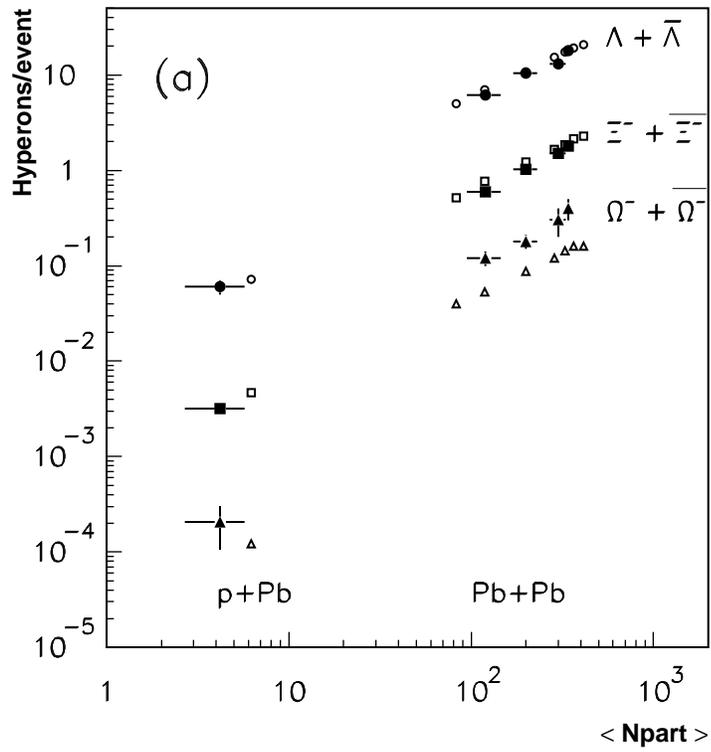
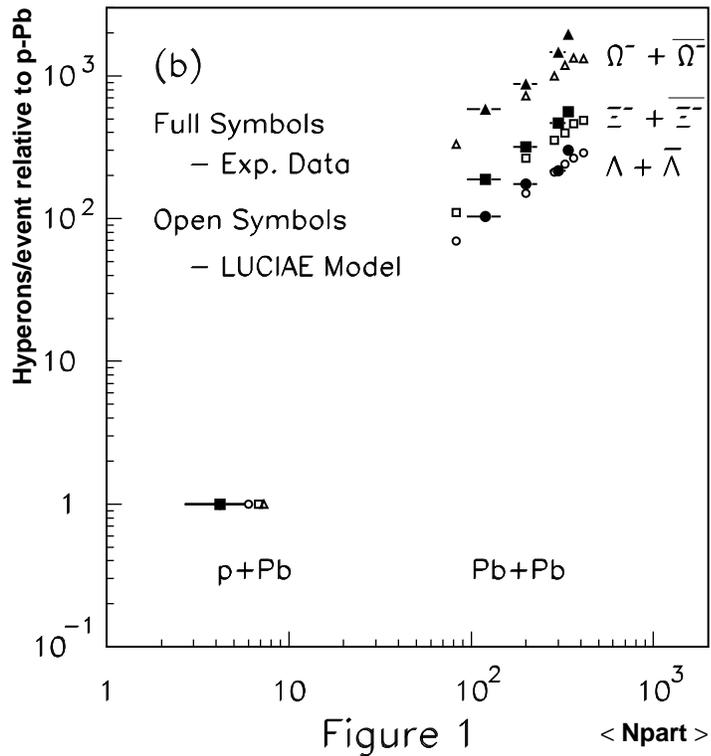
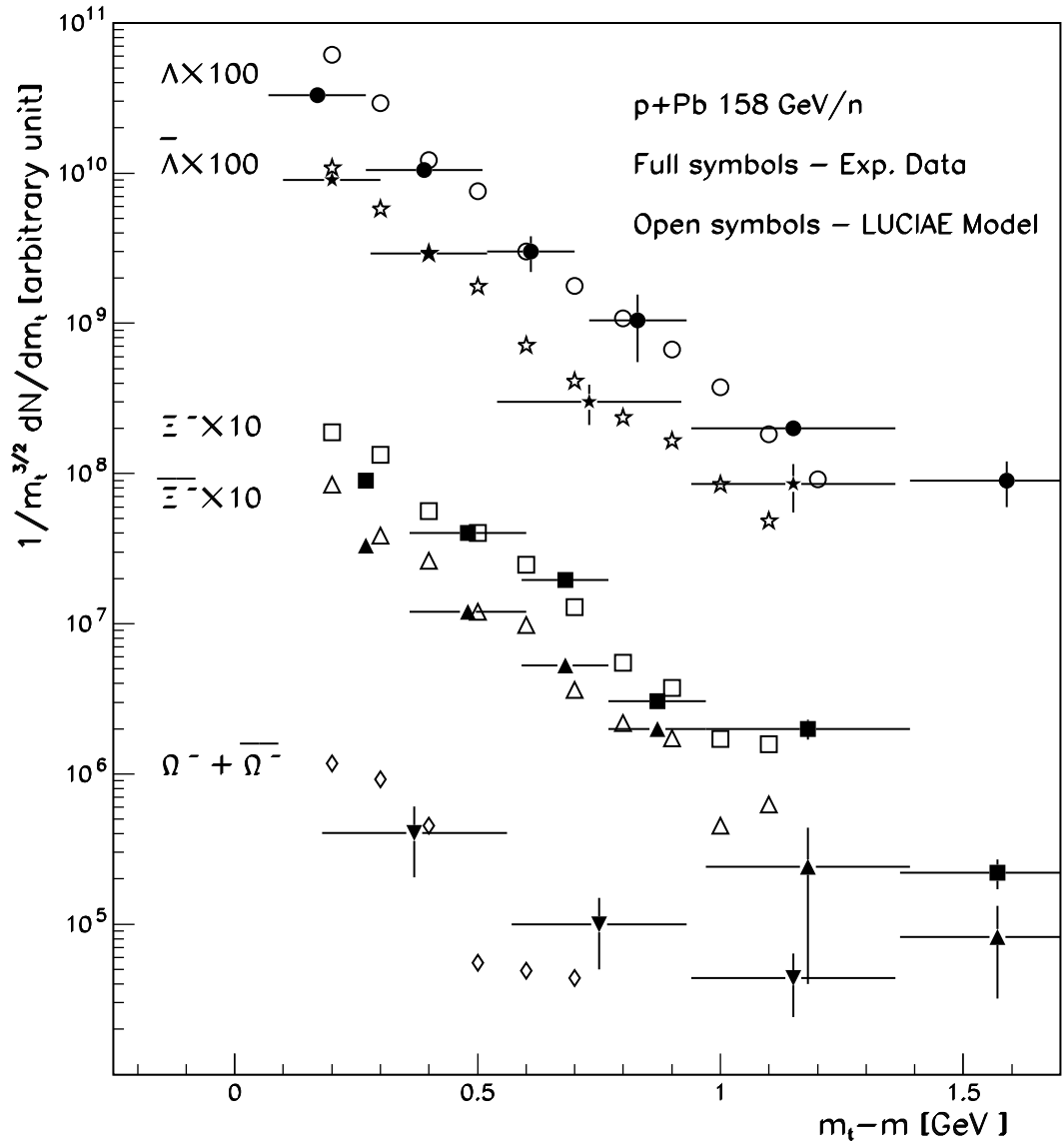
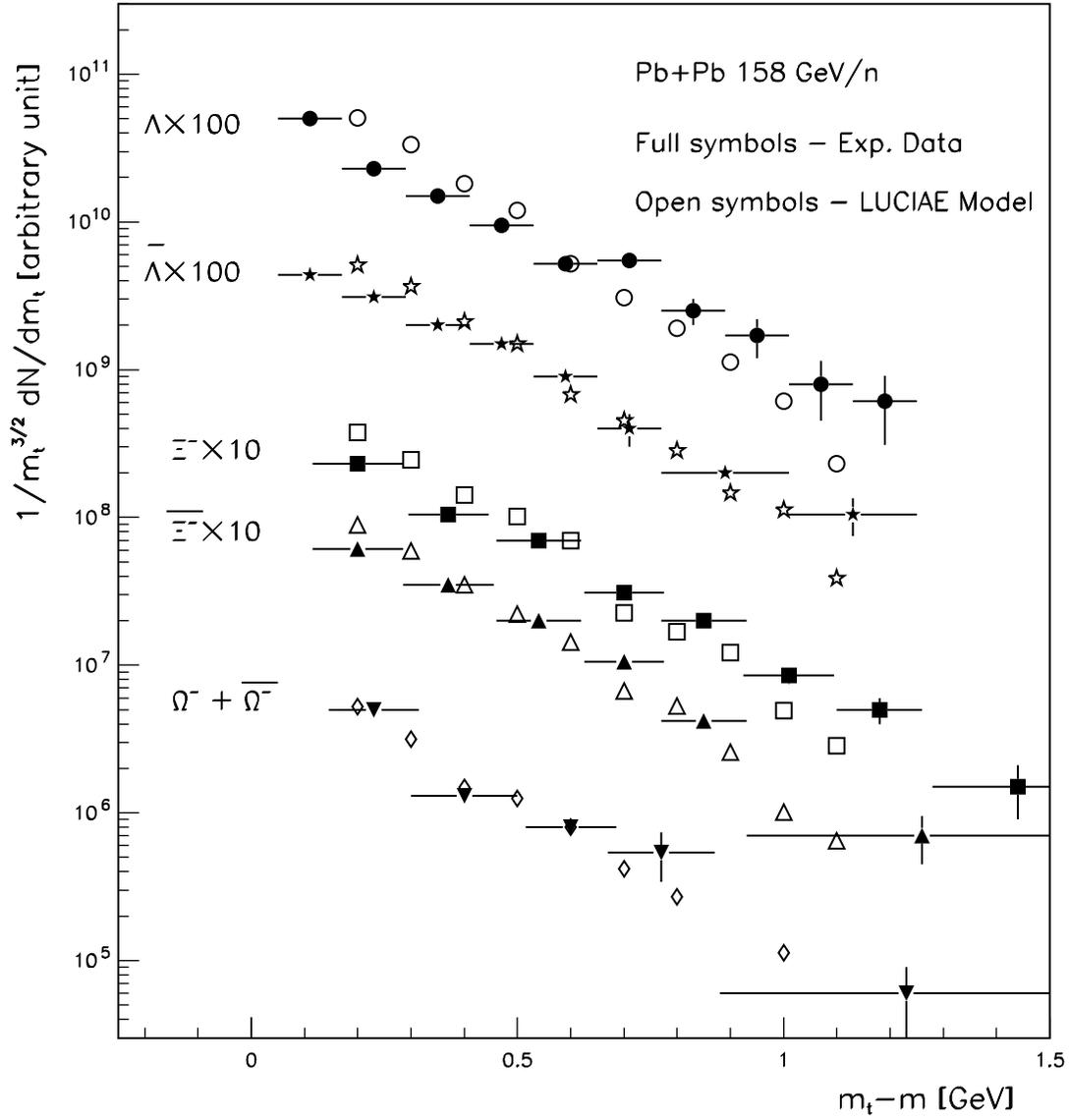
\end{document}